\PassOptionsToPackage{unicode}{hyperref}
\PassOptionsToPackage{hyphens}{url}
\documentclass[
]{article}
\usepackage{lmodern}
\usepackage{amssymb,amsmath}
\usepackage{ifxetex,ifluatex}
\ifnum 0\ifxetex 1\fi\ifluatex 1\fi=0 
  \usepackage[T1]{fontenc}
  \usepackage[utf8]{inputenc}
  \usepackage{textcomp} 
\else 
  \usepackage{unicode-math}
  \defaultfontfeatures{Scale=MatchLowercase}
  \defaultfontfeatures[\rmfamily]{Ligatures=TeX,Scale=1}
\fi
\IfFileExists{upquote.sty}{\usepackage{upquote}}{}
\IfFileExists{microtype.sty}{
  \usepackage[]{microtype}
  \UseMicrotypeSet[protrusion]{basicmath} 
}{}
\makeatletter
\@ifundefined{KOMAClassName}{
  \IfFileExists{parskip.sty}{%
    \usepackage{parskip}
  }{
    \setlength{\parindent}{0pt}
    \setlength{\parskip}{6pt plus 2pt minus 1pt}}
}{
  \KOMAoptions{parskip=half}}
\makeatother
\usepackage{xcolor}
\IfFileExists{xurl.sty}{\usepackage{xurl}}{} 
\IfFileExists{bookmark.sty}{\usepackage{bookmark}}{\usepackage{hyperref}}
\hypersetup{
  hidelinks,
  pdfcreator={LaTeX via pandoc}}
\hypersetup{colorlinks=true}

\author{}
\date{}

\begin{document}

\hypertarget{comment-on-the-statistics-wars-and-intellectual-conflicts-of-interest-by-d.-mayo}{%
\section{Comment on ``The statistics wars and intellectual conflicts of
interest'' by D.
Mayo}\label{comment-on-the-statistics-wars-and-intellectual-conflicts-of-interest-by-d.-mayo}}

\hypertarget{philip-b.-stark-department-of-statistics-university-of-california-berkeley}{%
\subsection{Philip B. Stark, Department of Statistics, University of
California,
Berkeley}\label{philip-b.-stark-department-of-statistics-university-of-california-berkeley}}

\hypertarget{april-2022}{%
\subsubsection{13 April 2022}\label{april-2022}}

\hypertarget{to-appear-in-conservation-biology}{%
\subsubsection{\texorpdfstring{To appear in \emph{Conservation
Biology}}{To appear in Conservation Biology}}\label{to-appear-in-conservation-biology}}

I enjoyed Prof.~Mayo's comment in \emph{Conservation Biology} (Mayo,
2021) very much, and agree enthusiastically with most of it. Here are my
key takeaways and reflections.

I agree with Prof.~Mayo that error probabilities (or error rates) are
essential to consider: if you don't give thought to what the data would
be like on the assumption that your theory is false, you are likely
reinforcing confirmation bias rather than establishing the validity of
your theory. As Mayo argues, ``stress-testing'' models and results is
crucial.

I also agree with Mayo that banning ``bright lines'' is not helpful.
Some applications really require a decision to be made or a conclusion
to be drawn. Should a species get specific legal protection, or not?
Should a pipeline be built through a particular area, or not? Should
drilling be permitted in a particular area, or not? Should there be
economic incentives for an agricultural management intervention that is
purported to sequester carbon, or not? Should a jurisdiction prohibit
harvesting wild foods from urban ecosystems for safety concerns, or not?
While such decisions are not made \emph{solely} on the basis of
\(P\)-values, banning bright lines may preclude making decisions in a
principled, reproducible way that controls error probabilities.

Conversely, no threshold for significance, no matter how small, suffices
to prove an empirical claim. As Fisher wrote:

\begin{quote}
{[}N{]}o isolated experiment, however significant in itself, can suffice
for the experimental demonstration of any natural phenomenon; for the
``one chance in a million'' will undoubtedly occur, with no less and no
more than, its appropriate frequency, however surprised we may be that
it should occur to \emph{us}. \textbf{In order to assert that a natural
phenomenon is experimentally demonstrable we need, not an isolated
record, but a reliable method of procedure.} In relation to the test of
significance, we may say that a phenomenon is experimentally
demonstrable when we know how to conduct an experiment which will rarely
fail to give us a statistically significant result. (Fisher, 1935,
emphasis added.)
\end{quote}

In replication lies empirical truth. While ``strict'' replication may be
impossible in ecology---``you cannot step in the same river twice''
(Heraclitus)---experiments, surveys, and other data collection campaigns
can be repeated, at least approximately. If minor changes in
circumstances alter the conclusions unpredictably, arguably the findings
are ``scientific observations'' but perhaps not ``biology'': the domain
to which findings generalize determines the scientific discipline (see,
e.g., Stark, 2018). If a result does not generalize to any other time or
place, it is not clear that it has scientific utility.

I also agree with Prof.~Mayo's thesis that abandoning \(P\)-values
exacerbates moral hazard for journal editors, although there has always
been moral hazard in the gatekeeping function. Absent any objective
assessment of the agreement between the data and competing theories,
publication decisions may be even more subject to cronyism, ``taste,''
confirmation bias, etc.

Throwing away \(P\)-values because many practitioners don't know how to
use them is like banning scalpels because most people don't know how to
perform surgery. Those who would perform surgery should be trained in
the proper use of scalpels, and those who would use statistics should be
trained in the proper use of \(P\)-values.

In my opinion, the main problems with \(P\)-values are: faulty
interpretation, even of genuine \(P\)-values; use of \emph{nominal}
\(P\)-values that are not \emph{genuine} \(P\)-values; and perhaps most
importantly, testing statistical hypotheses that have no connection to
the scientific hypotheses (see below).

A \(P\)-value is the observed value of any statistic whose probability
distribution is dominated by the uniform distribution when the null is
true. That is, a \(P\)-value is the observed value of any measurable
function \(T\) of the data that doesn't depend on any unknown
parameters, and for which---if the null hypothesis is
true---\(\Pr \{T \le x\} \le x\) for all \(x \in [0, 1]\). Reported
``nominal'' \(P\)-values often do not have that \textbf{defining}
property.

One reason a nominal \(P\)-value may not be a genuine \(P\)-value is
that calculating \(T\) may involve many steps, including data selection,
model selection, test selection, and selective reporting. Practitioners
often ignore all but the final step in calculating the nominal
\(P\)-value. That is, in reality, \(T\) is generally the composition of
many functions
\(T_n \circ T_{n-1} \circ \cdots \circ T_2 \circ T_1 (\cdot)\), but
often only the final step \(T_n(\cdot)\) is considered in calculating
the nominal \(P\)-value. If what is done to the data involves selection,
conditioning, cherry-picking, multiple testing, stopping when things
look good, or similar things, they all need to be accounted for, or the
result will not be a genuine \(P\)-value.

In my experience, perhaps the most pernicious errors in the use of
\(P\)-values in applications are Type III errors: \emph{answering the
wrong question}\footnote{There are other definitions of Type III errors,
  such as ``correctly rejecting the null, but for the wrong reason.''}
by testing a statistical null hypothesis that has nothing to do with the
scientific hypothesis, aside from having some words in common. A
statistical null hypothesis needs to capture the science, or testing it
sheds no light on the matter. For example, consider a randomized
controlled trial with a binary treatment and a binary outcome. The
\emph{scientific} null is that the treatment does not have an effect
(either subject by subject, or on average across the subjects in the
trial). A typical \emph{statistical} null is that the responses to
treatment and placebo are all IID \(N(\mu, \sigma^2)\). The scientific
null does not involve independence, normal distributions, or equality of
variances. A genuine \(P\)-value for the statistical null does not say
much about the scientific null: it is the chance that the difference in
observed means would be as large or larger than observed \emph{if the
responses were independent random samples from the same normal
distribution}.

Here is a more nuanced example: do wind turbines with longer rotors kill
more raptors than turbines with shorter rotors, per kW of generating
capacity? An analyst might posit a model for raptor-turbine collisions,
say a zero-inflated Poisson regression model with coefficients for rotor
length, peak RPM, capacity, some site characteristics, average and peak
windspeed, season, mean atmospheric visibility, the estimated size of
the relevant raptor population, and other covariates. The statistical
hypothesis is then that the coefficient of rotor length in the model is
zero. Even if one computes a genuine \(P\)-value for that statistical
hypothesis, what does it have to do with the original scientific
question? Why is a Poisson model appropriate? (For example, the Poisson
model assumes that collisions are independent, but some raptors migrate
in flocks.) Even if a Poisson model were appropriate, why should the log
of the rate of the Poisson distribution depend linearly (or in any other
parametric way) on those covariates, and not on anything else? The
\(P\)-value is the chance that the estimated coefficient would be as
large as it was \emph{if collisions followed a Poisson model with the
specified dependence of log rate on exactly those covariates, and the
true coefficient of rotor length in that model were zero}.

I close with a comment regarding likelihood-based tests, which are
mentioned briefly in Prof.~Mayo's commentary. Here, I disagree with her.
There are indeed tests that depend only on likelihoods or likelihood
ratios---and that allow optional stopping when ``the data look
good''---but that nonetheless rigorously control the probability of a
Type I error. Wald's sequential probability ratio test (Wald, 1947) is
the seminal example, but there are a host of other martingale-based
methods that give the same protections and give ``anytime-valid''
\(P\)-values.

\hypertarget{references}{%
\subsubsection{References}\label{references}}

Fisher, R.A., 1935 (1971). \emph{The Design of Experiments}, 9th
edition, MacMillan.

Mayo, D., 2022. The statistics wars and intellectual conflicts of
interest, \emph{Conservation Biology, 36,} e13861.

Stark, P.B., 2018. Before reproducibility must come preproducibility,
\emph{Nature, 557,} 613.

Wald, A., 1947 (1973). \emph{Sequential Analysis}, Dover, NY.

\end{document}